\documentclass[debug,overfull]{epl}

\usepackage{graphicx}
\usepackage{dcolumn}
\usepackage{bm}

\title{Modification to the pre-factor of the semiclassical propagator}
\author{Quanlin Jie\inst{1}\thanks{E-mail: qljie@whu.edu.cn},
Bambi Hu\inst{2}, \and Baowen Li\inst{3}}

\institute{
\inst{1}Department of Physics, Wuhan University,
Wuhan 430072, P. R. China\\
\inst{2}Department of Physics and Centre for Nonlinear Studies, Hong
Kong Baptist University, Hong Kong, China\\
\inst{3}Department of Physics, National University of Singapore,
117542 Singapore}

\pacs{03.65.Sq}{}
\pacs{02.70.-c}{}
\pacs{31.15.Ar}{}

\begin{document}
\maketitle

\date{\today}

\begin{abstract} 
We modify the pre-factor of the semiclassical propagator to improve its
efficiency in practical implementations. The new pre-factor represents the
smooth portion of an orbit's contribution, and leads to fast convergence in
numerical calculations. As an illustration of the accuracy and efficiency of
the resultant propagator, we numerically calculate overlaps between quantum and
semiclassical wave functions, as well as low-lying spectrum density in a
10-dimensional system contains unstable classical orbits.  This sheds
light on applying semiclassical propagator to high dimensional systems.  
\end{abstract}




Semiclassical propagator connects quantum dynamics with classical orbits. It
enables one to evaluate quantum quantities from classical orbits.  This is in
principle applicable to high dimensional quantum systems as an alternative
method to do first principle calculations.  Since Gutzwiller adds Maslov phase
to the Van Vleck semiclassical propagator~\cite{1}, many tests confirm that the
semiclassical propagator has remarkable accuracy~\cite{2,3,4,5}.  Another
appealing feature of the semiclassical method is easy to implement parallel
computation by requiring each node to handle some orbits. The final result is
simply a summation of each orbit's contribution. 




The fundamental approximation of semiclassical method is to treat the
Hamiltonian locally as a quadratic function of coordinates and momenta. By
expanding the Hamiltonian up to the second order around a phase space point,
each orbit is associated to a time dependent quadratic quantum Hamiltonian in
semiclassical calculations. In this sense, if numerical performance is not a
concern, all kinds of semiclassical formulations are as accurate as the Van
Vleck - Gutzwiller (VVG) propagator~\cite{1}.  The later comes from the
stationary phase approximation to the Feynman's path integral representation of
the quantum propagator.  In numerical calculations, however, the Herman and
Kluk's (HK) formulation of semiclassical propagator~\cite{6} is often a
favorite one\cite{7,8,9,10,11,12,12a,12b,12c}.


Like other formulations, the HK propagator suffers from the sign problem, i.e,
fluctuations in nearby orbits' contributions.  An orbit's contribution is
usually a complex number with a phase related to the classical action of the
corresponding orbit. The action changes rapidly from orbit to orbit. At the
same time, the pre-factor of an orbit's contribution increases with time
rapidly in the HK formulation, this makes the fluctuation increasing
drastically with time, and it smears out any useful information sooner or
later. The situation is especially worse for  systems containing
unstable orbits.




A remedy to the sign problem is to use block method to smooth out the
fluctuations~\cite{13,14,15,16,17,17a,17b}. The basic idea is same as the
standard method to do numerical integration of a fluctuating function.  Such
treatment usually results in an exponential damping term that is equivalent to
filter out contribution from unstable orbits. Another consequence of such 
treatment is that the formulation becomes complicated and the calculation of
each orbit's contribution is numerically more costly.

In this Letter, we present an approach to the sign problem by extracting smooth
portion of an orbit's contribution.  The resultant semiclassical propagator
converges much faster while maintaining its original accuracy. We achieve this
by modifying the pre-factor of the HK formulation with a positive factor.
%
%
The semiclassical evolution operator with new pre-factor reads
\begin{equation}\label{eq:1}
\hat{\mathbf{U}}_{sc}(t)=\int\frac{dz}{(2\pi\hbar)^{N}}
R_{z,t}\Delta_{z,t}e^{iS(z,t)/\hbar} \left|z_{t},\lambda\right\rangle 
\left\langle z,\lambda\right|.
\end{equation}
Here $z=(q,p)$ stands for a $2\times N$ dimensional phase space point with $q$
and $p$ being the coordinate and momentum respectively. An initial phase space
point $z$ arrives $z_t=(q_t,p_t)$ at time $t$ along classical orbit.  The
action of the orbit is $S(z,t)=\int_{0}^{t}(p\dot{q}-H)dt$ with $H$ being the
Hamiltonian of the system. The states $\left|z,\lambda\right\rangle$ and
$\left|z_{t},\lambda\right\rangle$ are Gaussian wave packets with $\lambda$ the
squeeze parameter. In coordinate representation, it is
\begin{equation}\label{eq:2}
\left\langle x|z,\lambda\right\rangle=
\left(\frac{\lambda}{\pi\hbar}\right)^{N/4}
\exp[-\frac{\lambda}{2\hbar}(x-q)^{2}+\frac{i}{\hbar}p(x-q)].
\end{equation}
The pre-factor $R_{z,t}$ is the square root of a determinant of the complex
stability matrix,
\begin{equation}\label{eq:3}
R_{z,t} = \sqrt{\det\left[\frac{1}{2}\left( M_{qq}+M_{pp}-i\lambda
M_{qp}+\frac{i}{\lambda}M_{pq} \right)\right] }, 
\end{equation}
where the $N\times N$ matrices $M_{qq}=\partial_q q_t(q,p)$, 
$M_{pp}=\partial_p p_t(q,p)$, $M_{qp}=\partial_p q_t(q,p)$, and 
$M_{pq}=\partial_q p_t(q,p)$
constitute the stability matrix $M=\partial_z z_t(z)$ of the orbit.



Without the modification factor $\Delta_{z,t}$, Eq. (\ref{eq:1}) is just the HK
propagator. Note that the role of squeeze parameter $\lambda$ is equivalent to
a scale transformation~\cite{16}, i.e., one obtains the same result by applying
the canonical transformation $q\rightarrow\sqrt{\lambda}q$, $p\rightarrow
p/\sqrt{\lambda}$ to the underlying classical dynamics while keeping
$\lambda=1$ in (\ref{eq:1}) to (\ref{eq:3}). In the following discussions, we
set $\lambda=1$. With this setting, the modification factor has a simple form,
\begin{equation}\label{eq:4}
\Delta_{z,t}=(2\beta+1)^{N}[\det(2\beta M^{T}M+1)]^{-1/2},
\end{equation}
where $\beta$ is a large positive number, and the superscript $T$ stands for
matrix transpose. Since the determinant of the symplectic stability matrix $M$
is unit, $\det M = 1$, and the inverse of $M$ is already known, we use
$\Delta_{z,t}=(2\beta+1)^{N}[\det(2\beta M^{T}+M^{-1})]^{-1/2}$ for practical
calculations. Computation cost of the determinant of a $2N \times 2N$ real
positive matrix is about the same as that of a $N \times N$ complex matrix,
thus calculation of the modification factor needs about the same computer
operations as that of the original pre-factor. This ensures the efficiency to
implement the semiclassical evolution operator (\ref{eq:1}).


The modification factor (\ref{eq:4}) represents the smooth portion of an
orbit's contribution. We obtain it from a kind of block treatment to the HK
propagator. The starting point is similar to that in Refs.
\cite{13,14,15,16,17}. It is also referred as Filinov transformation to the HK
propagator. The basic idea of this block approach is to replace one orbit's
contribution with a weighted average of nearby orbits' contribution. For weight
function in Gaussian form, the summation of the nearby orbits' contribution has
analytic result.  Here, the Hamiltonian within each block is approximated as a
time dependent quadratic function of coordinates and momenta associated with a
representing orbit. This is equivalent to treat the nearby orbits within a
block via linearized dynamics associated with the representing orbit. It is
apparent that the form of the weight function determines the performance of the
resultant propagator.


We perform the weighted average in coherent state (Gaussian wave packet)
representation of HK propagator, $\langle
z'_0,\lambda'|\hat{\mathbf{U}}^{HK}_{sc}(t)|z_0,\lambda'\rangle$.  Here
$\hat{\mathbf{U}}^{HK}_{sc}(t)$ is the semiclassical evolution operator
(\ref{eq:1}) without the modification term $\Delta_{z,t}$.  In principle, any
representation should give the same result, provided that one chooses proper
weight function. The coherent state representation is over complete, it
suffices to consider the case that the initial Gaussian is the same as the
final one, $z'_0=z_0$.  The squeeze parameter $\lambda'$ can be chosen
arbitrary. We set $\lambda'=1$, i.e. equal to $\lambda$ in Eq.  (\ref{eq:3}).
Under this specification, we choose a weight function in the following Gaussian
form:
\begin{equation}\label{eq:5}
W_\beta(z-z')=A_W\exp[-\frac{\beta}{2\hbar}(z_t - z_t')^2
 +i\epsilon(z_t - z_t')].
\end{equation} 
Here $A_W$ is the normalization constant. The phase points $z$ and $z'$ move to
$z_t$ and $z'_t$ along classical orbits at time $t$, respectively.  The width
parameter $\beta>0$ is a large number to ensure that the weight function is
effectively distributed near its center point. The imaginary part of the
exponent is a small adjusted phase to cancel fluctuation in the average
procedure. We choose $\epsilon$ to make the orbit starting from $z$ being a
stationary phase orbit.

After the weighted average, an orbit's contribution  to the propagator gains an
extra factor
\begin{equation}\label{eq:6}
\Delta_{z,t}= \left(\frac{2\beta+1}{\pi}\right)^N \int dz'
e^{-[2\beta(\delta z_{t})^{2}+(\delta z)^{2}+2(z-z_{0})\delta z]/4\hbar},
\end{equation}
where $\delta z=z'-z$, $\delta z_t=z'_t-z_t$, and $z$ is the initial phase
space point of the orbit. We obtain the above result by applying linearized
dynamics to the actions of the nearby orbits. Details of the procedure is
similar to that in Refs. \cite{13,14,15,16,17}.  Another approximation is that
the terms $(\delta z_t)^2+2(z_t-z_0)\delta z_t$, as well as a small
normalization term are dropped from the exponent of the integrand. These terms
are small compared to the first term of the exponent.

Note that the right hand side of (\ref{eq:6}) is real and positive definite,
i.e. the integrand's phase vanishes.  This comes from two facts: (1) Since we
set $\lambda'=\lambda=1$, the second order terms of an nearby orbit's action
cancel with the second order terms of the phase of the Gaussian wave packets;
(2) The first order terms of the action plus the first order terms of the phase
of the Gaussian wave packet are $(p_{t}-p_{0})\delta q_{t}-(q_{t}-q_{0})\delta
p_{t} -(p-p_{0})\delta q+(q-q_{0})\delta p$. Under linearized dynamics,
$(p-p_{0})\delta q-(q-q_{0})\delta p$ is symplectic, i.e., it is equal to
$(p_t-p_{0t})\delta q_t-(q_t-q_{0t})\delta p_t$, where $z_0=(q_0,p_0)$ moves to
$z_{0t}=(q_{0t},p_{0t})$ along classical orbit at time $t$. Thus the first
order terms of the integrand's phase are linearly dependent on $\delta
z_t=(\delta q_t,\delta p_t)$. By proper choice of the adjust parameter
$\epsilon$ in the weight function (\ref{eq:5}), the integrand's phase vanishes.

The integral in (\ref{eq:6}) represents an overlap of two classical density
distributions at time $t$. One of them is initially $\rho_1(0)= \exp\{-[(\delta
z)^{2}+2(z-z_{0})\delta z]/4\hbar\}$. After evolution for a period of time $t$
according to classical dynamics, its overlap with the density distribution
$\rho_2=\exp[-\beta (\delta z)^2/2\hbar]$ gives the integral in (\ref{eq:6}).

From consideration of classical dynamics, we argue that the second term in the
exponent of $\rho_1$ is negligible. At initial time $t=0$, the second density
distribution $\rho_2$ is highly localized in comparison with $\rho_1$. Within
the effective distributed area of $\rho_2$, $\rho_1$ is virtually a constant.
Thus $\rho_2$ dominates the integral and $\rho_1$ can be replaced by a constant
at initial time.  As time $t$ increases, phase space points stretch out in some
directions, and folding up in others.  Such stretching and folding of phase
space make the distribution $\rho_1$ distorted in phase space. The quadratic
term of $\rho_1$, i.e., $\rho_{11}=\exp[-(\delta z)^2/4\hbar]$ has only finite
effective distribution area that is distorted during evolution.  The distorted
area overlaps only partially with $\rho_2$ at time $t$. On the other hand, the
second term of $\rho_1$, i.e., $\rho_{12}=\exp[-(z-z_0)\delta z/2\hbar]$,
however, behaves in a quite different way.  This term is unbounded, and it is
smoother than the quadratic term. Thus the distortion of phase space does not
affect its overlap with the distribution $\rho_2$. Similar to the initial time
$t=0$, within the effective distribution area of $\rho_2$, $\rho_{12}$ is
effectively a constant, and can be dropped from the integral.

As usual, we treat the overlap between $\rho_{11}$ and $\rho_2$ at time $t$ via
linearized dynamics. The result is (\ref{eq:4}).  Since this modification
factor is independent of the wave packet $|z_0,\lambda'\rangle$, we obtain the
evolution operator in  the form of Eq. (\ref{eq:1}). This concludes our
theoretical derivation. 




We test the performance of the semiclassical evolution operator (\ref{eq:1})
via H\'enon-Heiles (HH) Hamiltonian~\cite{18}. This system contains unstable
classical orbits, and the original HK propagator converges only in a rather
short time scale. Properties of the HH system is widely investigated
classically, quantum mechanically, as well as semi-classically, including
improvement to the semiclassical propagator by block
treatments~\cite{14,15,16}. 



\begin{figure}[h]
\includegraphics[angle=-90,width=\columnwidth]{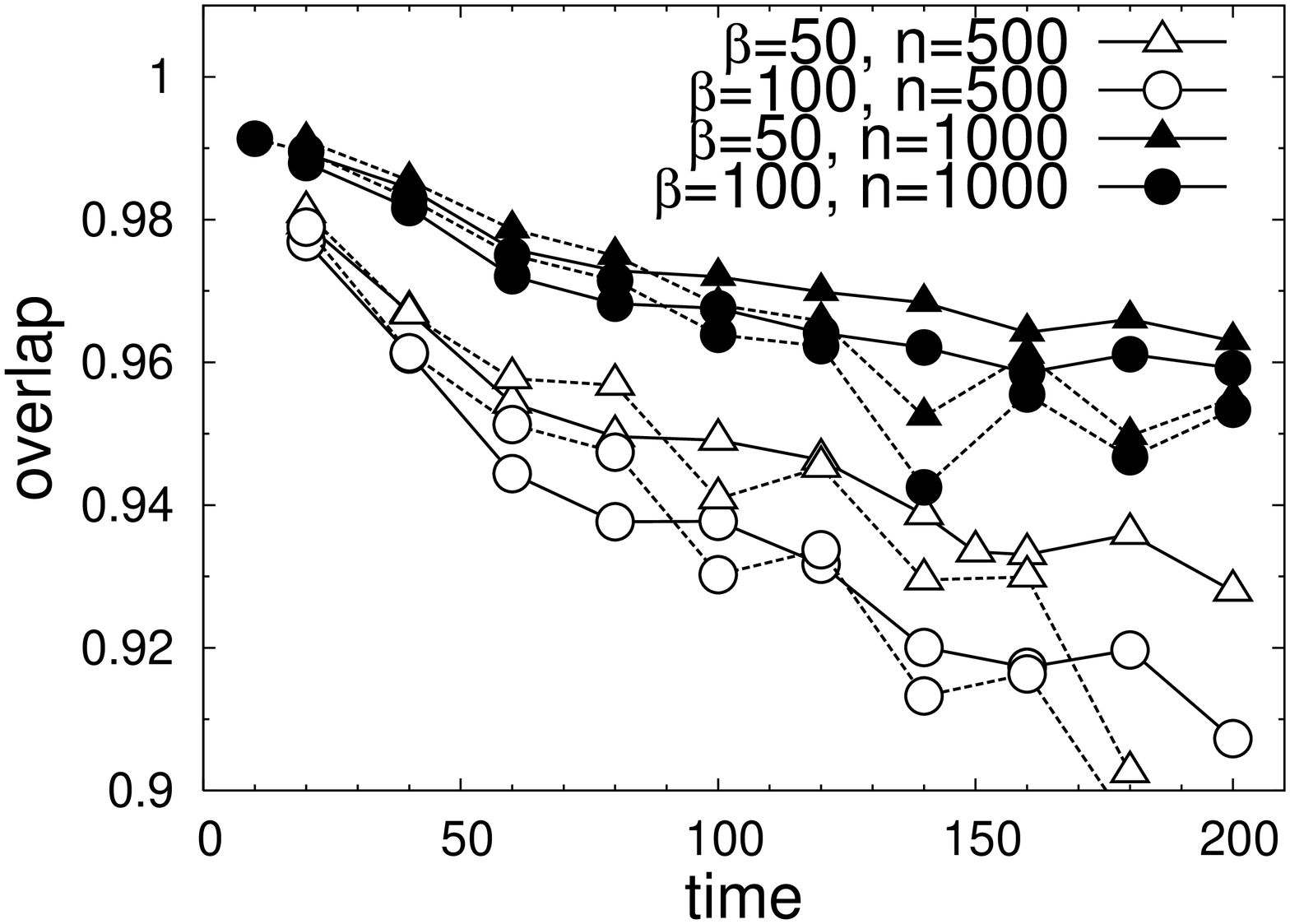}
\caption{
\label{fig1}
Overlap between normalized semiclassical wave function and exact quantum wave
function versus time. Solid and dashed lines are results of Eq. (\ref{eq:1})
and Herman's formulation~\cite{15}, respectively.  $\beta$ is the width
parameter, and $n$ is the number of orbits used in semiclassical calculations.
}
\end{figure}

We first compare semiclassical wave function with exact quantum result via
overlap between semiclassical and quantum wave functions.  Solid lines in
Fig. \ref{fig1} are the result of a particle of unit mass in the 2-dimensional
H\'enon-Heiles potential $v(x,y)=\frac12(\omega_x^2 x^2+\omega_y^2
y^2)+\lambda y(x^2+\eta y^2)$, where $\omega_x=1.3$, $\omega_y=0.7$,
$\lambda=-0.1$, $\eta=0.1$, and the Planck constant is set to $1$. The wave
function of the particle is initially a Gaussian with center position at
$(1,1)$ and center momentum vanished.  The semiclassical wave function is
normalized to unit before doing overlap.  These parameters are the same as
that of Ref.  \cite{15}, in which there is a similar calculation in a time
scale between $0$ to $100$, and it shows that the original HK propagator
converges only for time less than 50. We see that, within a normalization
constant, the semiclassical wave function possesses remarkable accuracy.  As
comparison, we show by dashed lines the corresponding result using formulation
of Ref.~\cite{15}, which is more suitable for calculation of wave function
than other existing formulations. It is evident that for time $t>100$,
Eq. (\ref{eq:1}) is more stable and converges faster.

As is shown in Fig. \ref{fig1}, the accuracy of the semiclassical wave function
is rather insensitive to the value of $\beta$, provided that $\beta$ is large
enough, or the effective distribution area of the weight function is
sufficiently localized.  Small value of $\beta$, or wider distributed weight
function, leads to faster convergence, i.e., one needs less orbits.  However,
if $\beta$ is improperly small, the linearized dynamics treatment to the
modification factor is invalid beyond certain time. When this happens, we find
that the normalization constant of the wave function becomes vanishing small
after certain time. We use this property to check the validness of the width
parameter $\beta$. 



\begin{figure}[h]
\includegraphics[angle=-90,width=\columnwidth]{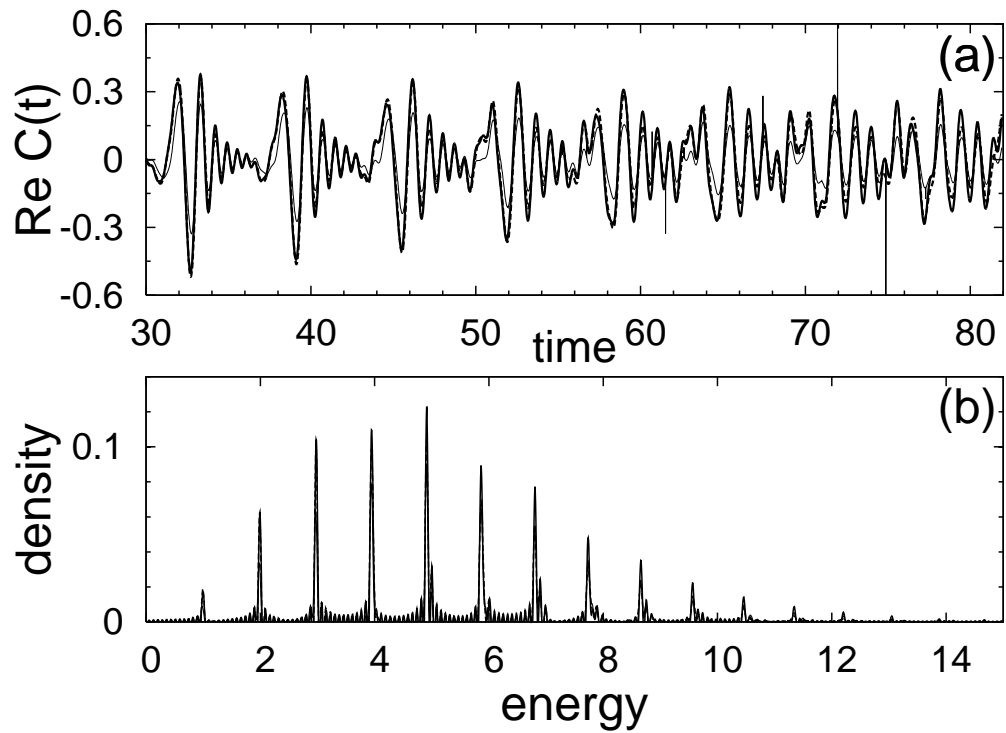}

\caption{
\label{fig2}
(a) Real part of the auto-correlation function $C(t)$ versus time.  (b) Real
part of the spectrum density versus energy.  Thick solid, dashed and thin
solid lines are exact quantum result and semiclassical result from Eq. (\ref{eq:1}),
as well as semiclassical result using formulation of Ref.~\cite{16},
respectively.
}
\end{figure}

Next, we test the semiclassical propagator's performance in finding energy
spectrum of quantum systems. In fact, calculation of energy spectrum in high
dimensional systems is fundamental important, and is still a challenge in many
systems.  Similar to Ref.~\cite{16}, we make numerical test in a high
dimensional H\'enon-Heiles system, i.e., a particle of unit mass moving in a
$N$-dimensional HH potential, $V(x)=\frac12\sum_{i=1}^N
x_i^2+\lambda\sum_{i=1}^{N-1}(x_i^2 x_{i+1}+ \eta x_{i+1}^3)$, where
$\lambda=0.1$, and $\eta=-0.3$.  The initial wave function is a Gaussian wave
packet whose central position is $q_i=2$ and central momentum vanishes,
$p_i=0$.  Classical orbit in such region is more unstable than that of Fig.
\ref{fig1}.  One can obtain energy spectrum from auto-correlation function by
Fourier transformation, or more advanced filter diagonalization~\cite{19}.  In
Fig. \ref{fig2}(a), we show the auto-correlation function versus time in a
2-dimensional case, and in Fig.\ref{fig2} (b) we show the real part of its
Fourier transformation, namely, the spectrum density. Here the thick solid
line is the exact quantum result, and the dashed line is semiclassical result
from Eq. (\ref{eq:1}).  The thin solid line is the semiclassical result using
formulation of Ref.~\cite{16}, an improved version of Walton's
formulation~\cite{14}.  We see that in a remarkable time scale, the
semiclassical auto-correlation function is very close to the quantum result.
Such time scale is enough to resolve the spectrum, and the semiclassical
spectrum density is able to reproduce quantum result very well.  This is
especially the case for the peaks of the spectrum density, whose positions
correspond to the energy spectrum.

The behavior of the auto-correlation function is about the same as that of the
wave function as shown in Fig. \ref{fig1}.  Our test shows that the spectrum
density converges faster than the wave function, and the peaks of the spectrum
density are not sensitive to the width parameter $\beta$.  In calculations of
Fig. \ref{fig2} (a), we use $10^4$ orbits by setting the width parameter to
$1,000$. However the spectrum density of Fig.  \ref{fig2} (b) is resultant of
$2,000$ orbits with the width parameter of $100$.  In fact, auto-correlation
function in shorter time is more important to determine the position of the
peaks of the spectrum density, and shorter time wave function converges faster
than that of longer time, and is more insensitive to the width parameter. This
explains the fast convergence rate of the spectrum density.  Note that,
although Eq. (\ref{eq:1}) gives more accurate correlation function in longer
times than that of Ref.~\cite{16}, 
the two semiclassical results in Fig. \ref{fig2} yield almost the same
spectrum density.  This property of the spectrum density sheds light on
applying semiclassical propagator to find energy spectrum in high dimensional
systems.

\begin{figure}[h]
\includegraphics[angle=-90,width=\columnwidth]{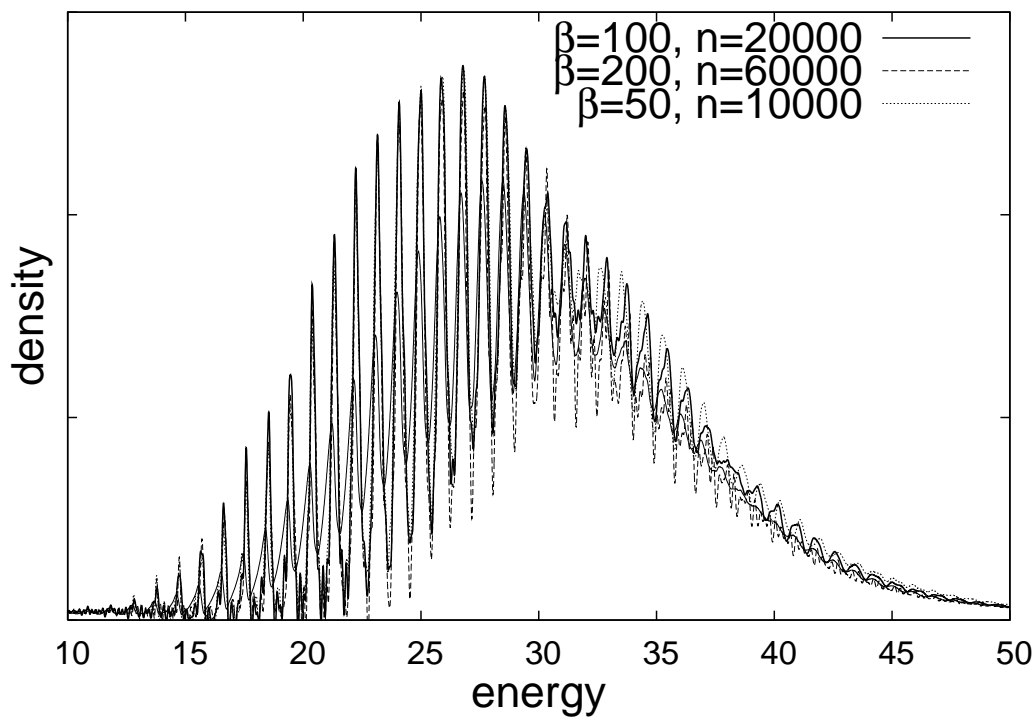}
\caption{
\label{fig3}
Spectrum density versus energy in 10-dimensional HH system. The number of
orbits $n$ and the width parameter $\beta$ in semiclassical calculations are
indicated accordingly. The thin solid line is the semiclassical result of
Ref.~\cite{16} with $\beta=100$ and $n=10000$.
}
\end{figure}

Our numerical calculation for high dimensional HH system confirms the above
properties in the lowlying energy.  The lowlying spectrum density converges
much faster than the higher ones, and the accuracy is quite insensitive to the
width parameter $\beta$. Figure \ref{fig3} is semiclassical spectrum density
versus energy in a 10-dimensional HH system.  To confirm the convergence, we
make calculations for different width parameters $\beta$.  The thick solid, dashed,
and dotted lines are for width parameter $\beta=100$, $200$, and $50$
respectively.  From Fig. \ref{fig3}, it is evident that the spectrum density
for energy less than 30 is well converged.  Small $\beta$ leads to fast
convergence without sacrificing the accuracy of lowlying spectrum.  We use
$20,000$, $60,000$, and $10,000$ orbits to produce the solid, dashed, and
dotted lines, respectively. In fact, for $\beta=50$, $2,000$ orbits is almost
enough to obtain the lowlying spectrum, i.e., the peak positions of the
spectrum density. Again, the thin solid line is the result using formulation
of Ref.~\cite{16}. Although it gives almost the same spectrum, result from
Eq. (\ref{eq:1}) has better resolution, which demonstrates better accuracy in 
longer times.








In summary, we have introduced a modification to the pre-factor of the
semiclassical propagator in the HK formulation.  It represents the smooth
portion of an orbit's contribution to the semiclassical propagator. Unlike
other formulations, this modification factor is just an overlap between two
classical density distributions. Our treatment to this classical overlap yields
a concise expression of the modification factor which is positive definite.
The resultant semiclassical propagator converges much faster while maintaining
its original accuracy.  This improvement to the convergence is prominent in the
calculation of lowlying spectrum density for high dimensional systems. This
result sheds light on the efforts towards applying semiclassical propagator as
a practical tool of first principle computation.


\bigskip

This work is supported in part by the National Natural Science Foundation
(Grant No. 10375042), the Research Fund of the State Education Ministry of
China, and the Research Fund of the Wuhan University.  BL is partly supported
by Faculty Research Grant of National University of Singapore.


\end{document}